\documentclass[sigconf, screen]{acmart}

\copyrightyear{2026}
\acmYear{2026}
\setcopyright{rightsretained}
\acmConference[ICSE '26]{2026 IEEE/ACM 48th International Conference on Software Engineering}{April 12--18, 2026}{Rio de Janeiro, Brazil}
\acmBooktitle{2026 IEEE/ACM 48th International Conference on Software Engineering (ICSE '26), April 12--18, 2026, Rio de Janeiro, Brazil}
\acmPrice{}
\acmDOI{10.1145/xxxxxxx.xxxxxx}
\acmISBN{xxx-x-xxxx-2025-3/26/04}

\usepackage[ruled,vlined,lined,commentsnumbered,linesnumbered]{algorithm2e}

\usepackage{algpseudocode}
\usepackage{amsmath}
\usepackage{amsfonts}
\usepackage{array}
\usepackage{arydshln}
\usepackage{balance}
\usepackage{booktabs}
\usepackage{color}
\usepackage{colortbl}
\usepackage{csvsimple}
\usepackage{diagbox}
\usepackage{dialogue}
\usepackage{enumitem}
\usepackage{fancybox}
\usepackage{flushend}
\usepackage{fontenc}
\usepackage{fontawesome}
\usepackage{graphicx}
\usepackage[utf8]{inputenc}
\usepackage{listings}
\usepackage{longtable}
\usepackage{lscape}
\usepackage{makecell}
\usepackage{marvosym}
\usepackage{moreverb}
\usepackage{multicol}
\usepackage{multirow}
\usepackage{pifont}
\usepackage{rotating}
\usepackage{setspace}
\usepackage{soul}
\usepackage{subcaption}
\usepackage{tabularx}
\usepackage{textcomp}
\usepackage[most]{tcolorbox}
\usepackage{threeparttable}
\usepackage{tikz}
\usepackage{url}
\usepackage{wrapfig}
 
\usepackage{amssymb}
\newcommand{\toolname}{{\textsc{Atomizer}}\xspace}

\newcommand{\PDG}[1]{$\delta\text{-PDG}$}
\newcommand{\PDGs}[1]{$\delta\text{-PDGs}$}
\newcommand{\Flexeme}[1]{\textit{Flexeme}}
\newcommand{\HDGNN}[1]{\textit{HD-GNN}}
\newcommand{\SmartCommit}[1]{\textit{SmartCommit}}
\newcommand{\agentA}{\textit{Purifier} agent\xspace}
\newcommand{\agentB}{\textit{Profiler} agent\xspace}
\newcommand{\agentC}{\textit{Grouper} agent\xspace}
\newcommand{\agentD}{\textit{Reviewer} agent\xspace}

\definecolor{MyBlue}{HTML}{BCC6DD}

\def\BibTeX{{\rm B\kern-.05em{\sc i\kern-.025em b}\kern-.08em
    T\kern-.1667em\lower.7ex\hbox{E}\kern-.125emX}}
    
\newcommand{\prompt}[3]{
\refstepcounter{myboxcounter}
\begin{tcolorbox}[colback=gray!10, colframe=black!80,
width=\linewidth, arc=2mm, auto outer arc, title={{\small #1}}, label={prompt:#2}, center]
{
\begingroup
\small
\linespread{0.8}\selectfont
#3
\endgroup
}
\end{tcolorbox}
}
\newcounter{myboxcounter}

\newcommand{\find}[1]{
\begin{tcolorbox}[tile,fontupper=\small,size=fbox,boxsep=2.2mm,boxrule=0pt,top=0pt,bottom=0pt,borderline west={0.5mm}{0pt}{black!50!white},colback=black!5!white]
#1
\end{tcolorbox}
}
\usepackage{booktabs}
\usepackage[table,xcdraw]{xcolor} 
\usepackage{multirow}

\begin{document}

\title{\toolname: An LLM-based Collaborative Multi-Agent Framework for Intent-Driven Commit Untangling}

\author{Kangchen Zhu}
\orcid{0000-0003-3876-0665}
\authornote{Kangchen Zhu, Shangwen Wang and Xiaoguang Mao are also with the State Key Laboratory of Complex and Critical Software Environment.}
\affiliation{%
  \institution{College of Computer Science and Technology, National University of Defense Technology}
  \city{Changsha}
  \country{China}}
\email{zhukangchen18@nudt.edu.cn}

\author{Zhiliang Tian}
\orcid{0000-0002-8906-5198}
\authornote{Zhiliang Tian and Shangwen Wang are the corresponding authors.}
\affiliation{%
  \institution{College of Computer Science and Technology, National University of Defense Technology}
  \city{Changsha}
  \country{China}}
\email{tianzhiliang@nudt.edu.cn}

\author{Shangwen Wang}
\orcid{0000-0003-1469-2063}
\authornotemark[1]
\authornotemark[2]
\affiliation{%
  \institution{College of Computer Science and Technology, National University of Defense Technology}
  \city{Changsha}
  \country{China}}
\email{wangshangwen13@nudt.edu.cn}

\author{Mingyue Leng}
\orcid{0009-0009-9042-3448}
\affiliation{%
  \institution{College of Computer Science and Technology, National University of Defense Technology}
  \city{Changsha}
  \country{China}}
\email{lengmengyue23@nudt.edu.cn}

\author{Xiaoguang Mao}
\orcid{0000-0003-4204-7424}
\authornotemark[1]
\affiliation{%
  \institution{College of Computer Science and Technology, National University of Defense Technology}
  \city{Changsha}
  \country{China}}
\email{xgmao@nudt.edu.cn}

\begin{abstract}
Composite commits, which entangle multiple unrelated concerns, are prevalent in software development and significantly hinder program comprehension and maintenance. Existing automated untangling methods, particularly state-of-the-art graph clustering-based approaches, are fundamentally limited by two issues. (1) They over-rely on structural information, failing to grasp the crucial \textbf{semantic intent} behind changes, and (2) they operate as ``single-pass'' algorithms, lacking a \textbf{mechanism for the critical reflection and refinement} inherent in human review processes. To overcome these challenges, we introduce \toolname, a novel collaborative multi-agent framework for composite commit untangling. To address the semantic deficit, \toolname employs an \textbf{Intent-Oriented Chain-of-Thought (IO-CoT)} strategy, which prompts large language models (LLMs) to infer the intent of each code change according to both the structure and the semantic information of code. To overcome the limitations of ``single-pass'' grouping, we employ two agents to establish a \textbf{grouper-reviewer collaborative refinement loop}, which mirrors human review practices by iteratively refining groupings until all changes in a cluster share the same underlying semantic intent. Extensive experiments on two benchmark C\# and Java datasets demonstrate that \toolname significantly outperforms several representative baselines. On average, it surpasses the state-of-the-art graph-based methods by over 6.0\% on the C\# dataset and 5.5\% on the Java dataset. This superiority is particularly pronounced on complex commits, where \toolname's performance advantage widens to over 16\%.
\end{abstract}



\maketitle

\section{Introduction}

In collaborative software development, developers usually make code changes and commit the changes to the repositories. Ideally, commits in version control systems (VCS) should be ``atomic'', addressing a single concern (such as a feature implementation or a bug fix) to enhance history clarity, code comprehension, and review processes~\cite{rigby2012contemporary, tao2012software, wang2019cora}. However, developers often create ``composite commits'' that bundle multiple unrelated concerns into a single commit. Such composite commits account for as much as 11\% to 40\% of real-world repositories~\cite{herzig2013impact, herzig2016impact, nguyen2013filtering, tao2015partitioning, fan2024detect}.

Composite commits pose significant risks to software maintenance and vulnerability management. For example, a real composite commit from the Pulumi project (Commit ID: 3988) \footnote{\url{https://github.com/pulumi/pulumi/commit/398878de31e42f7ec4485eab1c665f9aeea4c98a}} spanned 24 files with 131 diff regions, combining bug fixes, new features, and support updates. This complexity leads to two major issues: \textbf{(1) Automated Security Tool Failure}: Vulnerability detection models may mislabel the commit, corrupting their training data by incorrectly linking unrelated changes. \textbf{(2) Manual Maintenance Challenges}: Reverting feature A would inadvertently reintroduce vulnerability B, making maintenance difficult and error-prone. These issues severely hinder code reviews, complicate repository analysis, and reduce the effectiveness of automated defect prediction and vulnerability detection \cite{barnett2015helping, kirinuki2016splitting, li2024cleanvul}. Consequently, the research community is actively developing automated methods to decompose composite commits into atomic units.

Existing automated untangling approaches have evolved through three main paradigms. Early \textit{heuristic rule-based} approaches relied on human-specified heuristic rules to infer relationships between changes~\cite{barnett2015helping, herzig2013impact, kirinuki2016splitting, wang2019cora}. Subsequently, \textit{feature-based} approaches improved upon this by defining machine-learnable features~\cite{dias2015untangling, yamashita2020changebeadsthreader}. 
More recently, \textit{graph clustering-based} approaches have become prominent, representing code changes within a graph and applying clustering algorithms to group them based on structural dependencies~\cite{parțachi2020flexeme, li2022utango, shen2021smartcommit}. While these \textit{graph clustering-based} approaches are promising, they suffer from two fundamental limitations.

{\bf First, these methods over-rely on structural information while failing to leverage semantics adequately.} They group changes based on code dependencies, often overlooking the developer’s \textbf{intent}, which refers to the specific goals and motivation behind the changes. These methods can erroneously group changes that are structurally linked but semantically unrelated. Conversely, they may also fail to group changes that share the same intent but lack explicit structural dependencies~\cite{fan2024detect}. While developer intent may be implicitly reflected in commit messages, relying on them is insufficient for robustly capturing semantics. Prior studies ~\cite{safwan2019decomposing, al2022developers, casillo2024towards} have shown that commit messages are often noisy, incomplete, or missing altogether~\cite{wang2019cora, nguyen2013filtering}, making them an unreliable source for semantic understanding.

{\bf Second, these methods typically follow a ``single-pass'' process, without refinement or a feedback mechanism.} Although existing methods, such as GNN-based approaches, may employ internal iterations, they ultimately lack a global review and correction mechanism for their final output. This is quite different from how human developers work~\cite{herzig2013impact, barnett2015helping}: developers usually reflect on our first attempt, spot mistakes, and make improvements through iteration. Without such a refinement mechanism, these systems fail to detect or fix errors in their initial output, which often results in suboptimal or even logically inconsistent groupings.

To overcome these challenges, we propose \toolname, a novel collaborative multi-agent framework for automated composite commit untangling. Specifically, to capture the structure information, we employ a \agentA to represent the code changes with minimal change subgraphs (MCSs), which models the code dependency relations.
To better capture semantic information and mitigate the over-reliance on structural signals, we introduce an \textbf{Intent-Oriented Chain-of-Thought (IO-CoT)} prompting strategy. IO-CoT guides large language models (LLMs) to perform step-by-step reasoning that mirrors how humans infer the underlying \textbf{intent} behind code changes. By explicitly reasoning about each change’s purpose, IO-CoT helps LLMs uncover their deeper semantic meaning.
To address the limitations of ``single-pass'' grouping, the framework introduces a \textbf{grouper-reviewer collaborative refinement loop} with a \agentC and a \agentD. It follows a human-like review processing to iteratively self-correct until achieving a sound partition.

To evaluate the effectiveness of \toolname, we conduct extensive experiments on two widely-used benchmark C\# and Java datasets. The results demonstrate that \toolname significantly outperforms state-of-the-art (SOTA) baselines, achieving a notable average accuracy of 57\% on the task of changed node prediction, which aims to correctly group code changes into their corresponding concerns.
Furthermore, this superiority becomes even more pronounced when evaluating complex commits with large graphs, where \toolname widens its performance gap against traditional graph-clustering methods.
This paper makes the following contributions:

\begin{itemize}[left=0pt]
    
    \item We identify and address two key limitations of existing commit untangling methods: an over-reliance on structural signals without adequately leveraging semantic intent, and the absence of a refinement loop for iterative correction.
    \item We propose \toolname, a multi-agent framework for composite commit untangling, which integrates two components: (1) an \textbf{Intent-Oriented Chain-of-Thought (IO-CoT)} prompting strategy that accurately infers developer intent, and (2) an iterative \textit{grouper-reviewer collaborative refinement loop} that improves grouping quality to produce high-quality atomic commits.
    
    \item We conduct a rigorous and comprehensive evaluation of \toolname on widely-used C\# and Java datasets. The results demonstrate that \toolname achieves state-of-the-art performance, significantly outperforming existing approaches, particularly on complex commits with large graphs.\footnote{Our package is publicly available at: \url{https://zenodo.org/records/17592234}.}
\end{itemize}

\section{Background and Related Work}
\subsection{Automated Composite Commit Untangling}
Research on automated commit untangling has evolved through three main paradigms: heuristic rule-based, feature-based, and graph clustering-based approaches.

{\bf Heuristic Rule-based Approaches.}
Early work in this area relied on manually defined heuristics to infer relationships between code changes. For instance, \textit{Herzig et al.}~\cite{herzig2013impact} decomposed changes into individual operations and used confidence voters to predict relationships between them. Similarly, \textit{Kirinuki et al.}~\cite{kirinuki2016splitting} proposed suggesting commit splits based on a database of historical changes, while \textit{CoRA}~\cite{wang2019cora} employed dependency analysis and PageRank to identify significant changes for reviewers. Another approach, \textit{ClusterChanges}~\cite{barnett2015helping}, organized modifications primarily based on def-use and use-use code relationships. While foundational, these methods are often labor-intensive to design and can lack generalizability.

{\bf Feature-based Approaches.}
To reduce the reliance on manual rule creation, subsequent methods focused on learning from features. \textit{EpiceaUntangler}~\cite{dias2015untangling} collected fine-grained features, such as whether changes occurred in the same package or method, and used a clustering algorithm to group them. Building on this, \textit{Yamashita et al.}~\cite{yamashita2020changebeadsthreader} introduced \textit{ChangeBeadsThreader} (CBT), an interactive environment that uses a simplified version of this feature-based clustering as an initial step, after which developers can manually refine the results by splitting or merging clusters. However, feature-based approaches are also limited by the predefined features and cannot achieve satisfactory untangling results for developers.

{\bf Graph Clustering-based Approaches.}
More recently, the approach has shifted towards representing commits as graphs and applying clustering algorithms. \textit{Flexeme}~\cite{parțachi2020flexeme} builds multi-version dependency graphs with name flows and applies Agglomerative Clustering. \textit{SmartCommit}~\cite{shen2021smartcommit} partitions Diff-Hunk Graphs via edge shrinking for interactive refinement. \textit{UTango}~\cite{li2022utango} clusters code change embeddings generated by a Graph Neural Network (GNN). \textit{HD-GNN}~\cite{fan2024detect} uses a hierarchical GNN to detect hidden dependencies across commits. Despite recent progress, graph-based methods still face two core limitations. First, they rely on structural signals while failing to capture the semantic intent behind code changes. This often leads to groupings that do not reflect the developer’s intent. Second, they often lack a refinement mechanism. Once an initial grouping is made, the system cannot revisit or correct it. These limitations restrict the effectiveness of existing methods and point to the need for more intent-aware and self-correcting solutions.

\subsection{Developer Intent Analysis in SE}
Understanding developer intent is crucial for enhancing the accuracy, relevance, and usability of automated software engineering tools, as it helps bridge the gap between low-level code changes and high-level development objectives. Intent modeling has become increasingly significant in tasks such as requirements traceability~\cite{hey2019indirect}, code generation~\cite{lahiri2022interactive}, comment generation~\cite{mu2023developer, geng2024large, zhang2024multi}, and binary summarization~\cite{zhu2025misum}.
A common strategy involves extracting intent from commit messages. For instance, some approaches classify sentences in commit messages into categories such as ``Decision'' and ``Rationale'' to build knowledge graphs~\cite{dhaouadi2023towards}, while others use pre-trained models to identify rationale-bearing sentences~\cite{casillo2024towards}. However, these methods heavily depend on the availability and quality of human-written messages, which are often sparse or ambiguous in practice~\cite{wang2019cora, nguyen2013filtering}. Despite these efforts, existing untangling techniques have yet to sufficiently integrate developer intent, leaving considerable room for improvement in aligning automated untangling results with the developers' actual goals.

\subsection{Large Language Models for Code Reasoning}

{\bf Large Language Models for Code.}
The application of Large Language Models (LLMs) has marked a significant paradigm shift in numerous software engineering tasks. This evolution began with foundational models like  \textit{CodeBERT}~\cite{feng2020codebert} and \textit{CodeT5}~\cite{wang2021codet5}, which demonstrated that pre-training on massive code corpora enables LLMs to capture the deep, contextual semantics of programs. The field then saw the rise of powerful, large-scale proprietary models, such as OpenAI's \textit{Codex}~\cite{wang2021codet5}, which powered the first generation of GitHub Copilot, and Google's \textit{AlphaCode}~\cite{li2022competition}, which showed competitive performance on programming challenges. More recently, the development of capable open-source models specifically tuned for code, including Meta's \textit{CodeLlama}~\cite{roziere2023code}, BigCode's \textit{StarCoder}~\cite{lozhkov2024starcoder}, and the powerful \textit{DeepSeekCoder}~\cite{guo2024deepseek}, has further democratized this capability. These modern models demonstrate a remarkable proficiency in comprehending the intricate semantics of programs, enabling them to understand the high-level purpose of code changes, a crucial prerequisite for the untangling task.

{\bf Chain-of-Thought for Code Reasoning.} 
To tackle complex reasoning, a key technique is Chain-of-Thought (CoT) prompting. CoT has proven exceptionally effective when adapted for code-specific reasoning tasks, as it guides the model to articulate its intermediate logical steps. For instance, researchers have successfully applied CoT to infer the purpose of a change for commit message generation~\cite{yang2024chain, zhang2024using}. Similarly, CoT has been used to enhance complex tasks such as code generation~\cite{wang2023review, mu2024clarifygpt},  code completion~\cite{li2023cctest, husein2024large}, and vulnerability detection \cite{lu2024grace, du2024vul}, all of which demand a deep understanding of code logic rather than just surface-level syntax. These successes underscore the promise of applying CoT-based reasoning to our problem: inferring the actual intent behind tangled code changes.

{\bf LLM-based Multi-Agent Systems for Code Tasks.} 
A Multi-Agent System (MAS) is a framework consisting of multiple interacting agents that collaboratively solve complex problems, which would be beyond the capability of a single agent. MAS has diverse applications in software engineering, including requirements engineering~\cite{ataei2024elicitron, jin2024mare, sami2024ai}, code generation~\cite{wang2023intervenor, zhang2024pair, zan2024codes, liu2024agents4plc}, and software maintenance~\cite{hu2023large, qin2025s}. MAS architecture often follows several patterns. Role-Playing frameworks, like \textit{MARE}~\cite{jin2024mare} and \textit{MetaGPT}~\cite{hong2023metagpt}, offer structured workflows by emulating established software development processes. Task Decomposition frameworks, such as \textit{CodeS}~\cite{zan2024codes} and \textit{MegaAgent}~\cite{wang2024megaagent}, manage complexity by breaking large problems into smaller, more manageable tasks. Additionally, Feedback-Driven frameworks, including \textit{INTERVENOR}~\cite{wang2023intervenor} and \textit{SpecRover}~\cite{ruan2024specrover}, improve output quality and robustness by enabling self-correction through iterative feedback. These architectural patterns highlight the potential of multi-agent systems to orchestrate complex reasoning and self-correction, positioning them as a promising solution for challenges such as commit untangling.

\section{Motivation}
\label{sec:motivation}

\begin{figure*}[!t]
  \centering
  \includegraphics[width=0.95\linewidth]{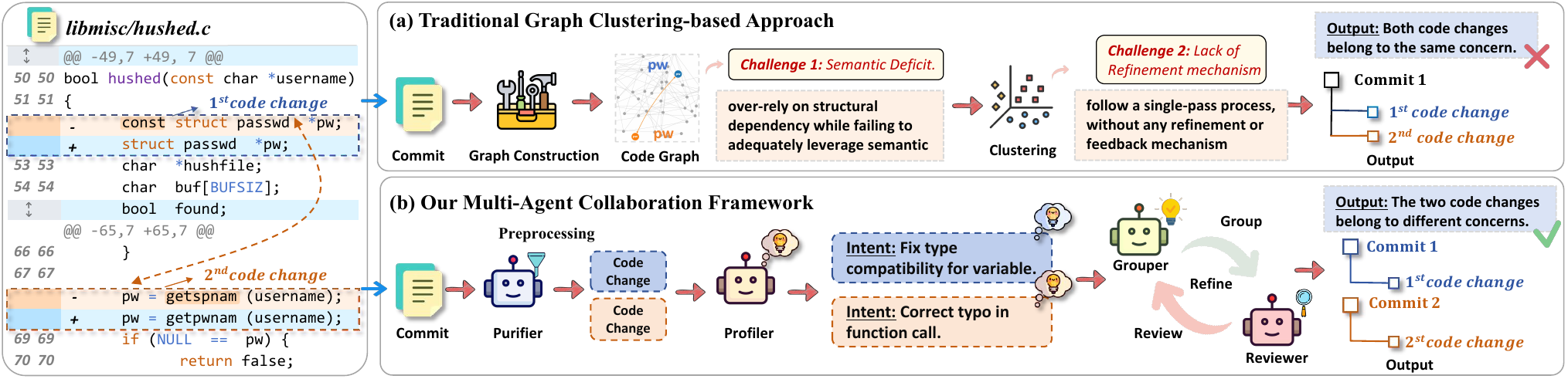}
  \caption{A motivating example using a real composite commit (ID: 2ba18ea) from the \textit{shadow-maint} project. The commit contains two changes with different rationales: a type fix (removing \texttt{const}) and a typo correction (\texttt{getspnam} → \texttt{getpwnam}).}
  \label{fig:motivation}
\end{figure*}

To illustrate the fundamental limitations of existing untangling methods, we use a real-world commit from the \textit{shadow-maint} project\footnote{\url{https://github.com/shadow-maint/shadow/commit/2ba18ea4a90fc1502c56032a63729b08cd13cc80}} as a motivating example. As shown in Figure~\ref{fig:motivation} (left), the commit contains two functionally distinct changes bundled together:
\begin{itemize}[left=0pt]
    \item Change 1 (Type Compatibility Fix): It resolves a compilation warning by removing the \texttt{const} qualifier from the \texttt{struct passwd *pw} variable declaration.
    \item Change 2 (Typo Correction): It fixes a typo by correcting a function call from \texttt{getspnam} to \texttt{getpwnam}.
\end{itemize}

\textbf{Challenge 1: Critical Semantic Deficit.}
Typically, existing methods tend to over-rely on structural information while failing to effectively incorporate semantic understanding. They typically group code changes based on syntactic dependencies, often overlooking the developer’s underlying intent—the purpose or rationale driving each change. As shown in Figure~\ref{fig:motivation}(a), although the two edits are structurally connected through a shared variable \texttt{pw}, they actually serve two unrelated purposes: one addresses a type fix, while the other corrects a typographical error. Due to the strong structural link, graph-based methods misinterpret the relationship and incorrectly cluster them into a single concern. This example highlights how the lack of semantic awareness in traditional approaches can lead to significant untangling errors.

\textbf{Challenge 2: Lack of a Refinement Mechanism.}
Another limitation of existing graph-based approaches is their nature as a ``single-pass'' process, which lacks iterative refinement. They produce a result without any mechanism for reflection or self-correction. Even if a system could infer the basic intents, an initial greedy grouping might still be flawed. For example, an automated agent might propose to group the changes because both are categorized as ``Fixes''. Without a critical review process, this plausible but incorrect grouping would become the final output. This is in stark contrast to human best practices, where review and iterative refinement are essential for quality.

To overcome the above challenges, our method \toolname adopts a multi-agent collaboration framework, as shown in Figure~\ref{fig:motivation}(b). \textbf{(1) To tackle the first challenge,} the \agentB analyzes each code change and infers its semantic intent, capturing not just what was changed, but why it was changed. For example, it identifies intents such as ``Fix: Correct type compatibility for variable'' or ``Fix: Correct typo in function call''. This intent-aware analysis allows \toolname to distinguish changes that are structurally connected but semantically unrelated, which traditional dependency-based methods often fail to do. \textbf{(2) To address the second challenge,} \toolname goes beyond just making an initial guess. Unlike previous approaches that terminate the processing after the first attempt, \toolname adopts a refinement loop: the initial grouping generated by \agentC is reviewed by \agentD, which plays the role of an expert reviewer; with a global and holistic view, \agentD evaluates whether the grouped changes actually belong together. In this example, it correctly recognizes that a ``type fix'' and a ``typo fix'' address different issues, and therefore would \texttt{REJECT} the incorrect grouping. This feedback then triggers a refinement cycle, ultimately leading to a correct and logically coherent result. This example demonstrates \toolname's superiority. It moves beyond brittle structural analysis by first inferring developer intent to understand the changes, and then, more importantly, by introducing a human-like collaborative review and refinement loop to ensure the final result is logically sound. This marks a paradigm shift from single-pass analysis to robust and self-correcting reasoning.

\section{Approach}
\label{sec:approach}

\begin{figure*}[!t]
  \centering
  \includegraphics[width=0.9\linewidth]{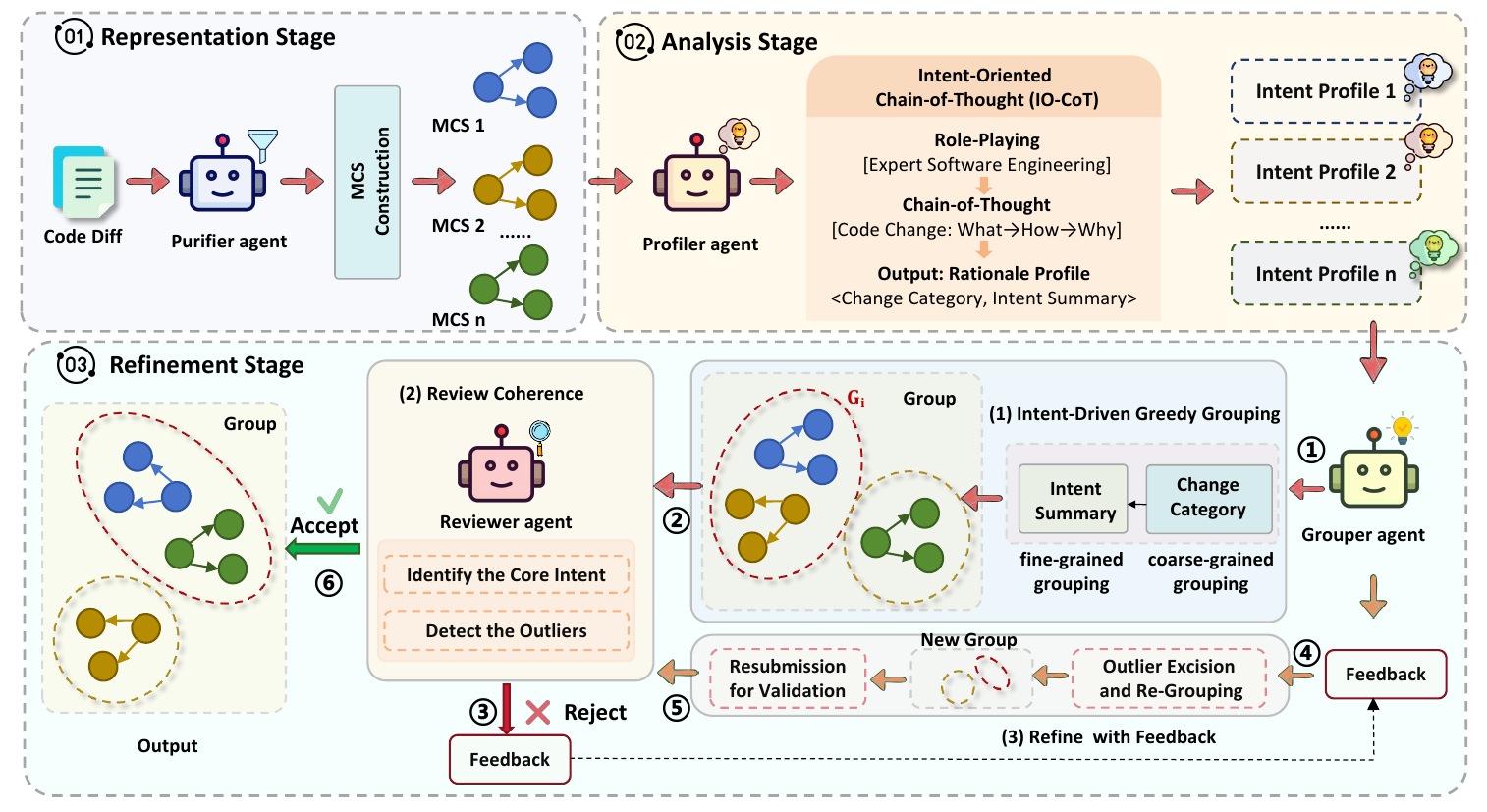} 
\caption{The overview of \toolname, which adopts a multi-agent framework for composite commit untangling. The \agentA first preprocesses the code diff into focused Minimal Change Subgraphs (MCSs). Subsequently, the \agentB employs an Intent-Oriented Chain-of-Thought (IO-CoT) strategy to infer the developer's intent for each MCS, addressing the semantic deficit. Finally, the \agentC and \agentD engage in a collaborative ``group-review-refine'' loop, enabling the framework to reflect, self-correct, and produce logically coherent atomic commits.}
  \label{fig:Overview of Atomizer.}
\end{figure*}

\subsection{Overview}
\label{sec:overview}
Figure~\ref{fig:motivation} illustrates an overview of \toolname, which employs a collaborative multi-agent framework that systematically untangles composite commits. To overcome two major limitations in traditional methods (semantic deficits and the lack of a refinement loop), \toolname first interprets the developer's intent behind code changes, then critically reviews and refines the groupings. The framework operates through a three-stage pipeline, with a single foundational LLM managing four specialized agents: \agentA, \agentB, \agentC, and \agentD. The three stages are as follows:
\begin{enumerate}[label=\arabic*., leftmargin=*]

    \item \textbf{Representation Stage:} The \agentA performs a preprocessing step to represent the code change information with \textit{Minimal Change Subgraphs} (MCSs). It parses a raw code file to obtain MCSs with structural information, which provide a clean and concise representation for the following stages.
    
    \item \textbf{Analysis Stage:} To address the \textbf{semantic deficit}, the \agentB performs a deep semantic analysis on each MCS: It leverages a novel \textit{Intent-Oriented Chain-of-Thought (IO-CoT)} prompting strategy to infer the developer's underlying intent according to the semantic information on code changes; and then it generates a structured \textbf{Intent Profile} for each change.
    
    \item \textbf{Refinement Stage:} To address the limitation of ``single-pass'' grouping, this stage simulates a realistic code review process involving both a \agentC and a \agentD. The \agentC first applies an \textit{intent-driven greedy grouping} algorithm to produce an initial grouping result. Next, the \agentD serves as a critical reviewer, evaluating the global logical coherence of the proposed grouping. Based on this evaluation, the \agentD provides feedback to the \agentC, who then revises the grouping accordingly. This creates an iterative \textbf{Grouper-Reviewer Collaborative Refinement Loop}, enabling the framework to progressively improve and converge toward an accurate grouping.

\end{enumerate}

\subsection{Preprocessing Stage: Minimal Change Subgraphs Construction}
\label{sec:agentA}

To improve the efficiency of analyzing code changes and eliminate interference from unrelated code statements, the \agentA constructs refined code subgraphs called \textit{Minimal Change Subgraphs} (MCSs). 
These subgraphs are designed to capture only the essential information about code changes while filtering out irrelevant structural noise. Each MCS is a subgraph of the original abstract syntax tree (AST), where nodes represent individual statement-level code elements (e.g., assignments, method calls, declarations) and edges represent syntactic relationships derived from the AST structure (e.g., parent-child relations, block scopes). Specifically, each MCS contains two key components:
\begin{itemize}[left=0pt]
    \item \textbf{Core Change Set:} A set of syntactically and logically interconnected statement-level changes (additions or deletions).
    \item \textbf{Essential Semantic Context:} A minimal set of unchanged statements (e.g., variable declarations, method signatures) essential for the semantic interpretation of the Core Change Set.
\end{itemize}
By focusing on these reduced, context-aware subgraphs, MCSs help isolate meaningful change patterns while suppressing noise from unrelated surrounding code. The \agentA constructs MCSs through a four-step process: 
\begin{enumerate}[label=(\arabic*), leftmargin=*]
\item \textbf{Seed Node Identification}: First, it parses the input difference using \textit{Tree-sitter}\footnote{\url{https://tree-sitter.github.io/tree-sitter/}} to map each modified line to its corresponding statement-level node in the AST. These identified nodes serve as the initial \textit{seed nodes} for the next analysis.

\item \textbf{Identifying Core Change Set through Intra-Change Dependency Analysis}: This step identifies the \textbf{Core Change Set}, which is a tightly coupled group of code modifications within the seed nodes. It analyzes dependency relationships only among the seed nodes, focusing on two types of dependencies: \textbf{data dependency}, where one changed statement uses variables or methods defined in another, and \textbf{control dependency}, where the execution of one changed statement depends on another, such as being nested in conditionals or loops. By linking seed nodes with these dependencies, the method clusters them into the logically coherent \textbf{Core Change Set}. Each set corresponds to a connected subgraph of modified AST nodes linked by internal data and control dependency edges.

\item \textbf{Extracting Essential Semantic Context via Bounded Backward Program Slicing}: This step extracts the \textbf{Essential Semantic Context} required to understand each \textbf{Core Change Set}. By performing \textit{bounded backward program slicing} on the AST, it traces backward from the modified nodes to gather a minimal set of unchanged statements, such as variable declarations and method signatures, which are essential for semantically interpreting the \textbf{Core Change Set}. The result is an enriched subgraph that combines the modified nodes with these crucial unchanged context nodes, where edges represent AST parent-child and control/data dependencies. 

\item \textbf{MCS Finalization}: By combining each \textbf{Core Change Set} with its \textbf{Essential Semantic Context}, the process produces the finalized Minimal Change Subgraphs (MCSs). 
\end{enumerate}

\subsection{Analysis Stage: Semantic Analysis via Intent-Oriented CoT}

\label{sec:agentB}
Next, to address the \textbf{semantic deficit} of prior work, the \agentB analyzes each MCS to infer the change intent. This is formalized as an \textbf{Intent Profile}, a structured data object containing:

\begin{itemize}[left=0pt]
    \item \textbf{Change Category:} A high-level classification of the change's purpose (e.g., \texttt{Bug Fix}, \texttt{Feature Addition}, \texttt{Refactoring}).
    \item \textbf{Intent Summary:} A concise summary suitable for a standard commit message.
\end{itemize}

\prompt{Prompt 1: Intent-Oriented Chain-of-Thought (IO-CoT)}{prompt:1}{
\label{prompt:IO-CoT}
You are an expert software engineer. Your objective is to meticulously analyze the following code change by following a structured reasoning process to infer its intent.

\vspace{0.1cm}
\hrule
\vspace{0.1cm}
\textbf{Input Code Change (Minimal Change Subgraph):}

\texttt{--- a/[file\_path]}

\vspace{0.1cm}

\texttt{+++ b/[file\_path]}

\vspace{0.1cm}

\texttt{[Code Diff Content]}

\vspace{0.1cm}
\hrule
\vspace{0.1cm}

Please perform your analysis by strictly following the \textbf{what -> how -> why} reasoning hierarchy. \textbf{Think step-by-step}:
\begin{itemize}[leftmargin=*, topsep=0pt, partopsep=0pt, itemsep=0pt, parsep=0pt]
    \item \textbf{Step 1: Literal Code Change Description (What):} Describe the exact code-level modifications. What was added, removed, or changed at the syntax level?
    \vspace{0.1cm}
    \item \textbf{Step 2: Functional Impact Analysis (How):} How does this change achieve its effect? What is the direct operational consequence on the program's behavior?
    \vspace{0.1cm}
    \item \textbf{Step 3: Change Category Inference (Why):} Why was this functional change necessary? What was the developer's ultimate goal? Choose one category and briefly justify your choice: [Bug Fix, Feature, Refactoring, Performance, Documentation, Test, Others].
    \vspace{0.1cm}
    \item \textbf{Step 4: Intent Summary Synthesis:} Based on your entire analysis (What, How, and Why), synthesize a single, concise summary written in the imperative mood (e.g., ``Fix...'', ``Add...'').
\end{itemize}

Please return the results in a structured JSON format.
}

To generate the intent profiles accurately, the \agentB employs a novel \textbf{Intent-Oriented Chain-of-Thought (IO-CoT)} prompting strategy on an LLM. IO-CoT inspires the LLM to replicate an expert developer's cognitive workflow (``what -> how -> why'') and requires a structured output, ensuring high-quality and consistent intent inference (see Prompt~\ref{prompt:IO-CoT} for details): 

\begin{itemize}[, leftmargin=*]

\item \textbf{Role-playing:} The prompt begins by assigning the LLM the role of an ``expert software engineer''. This activates its internal knowledge of software engineering, enabling it to produce outputs that are more professional in terminology, reasoning, and domain relevance.

\item \textbf{Chain-of-Thought:} IO-CoT is designed to inspire the LLM to replicate an expert developer's cognitive workflow, which systematically progresses from observing a change to understanding its intent. This workflow is structured into a clear ``what -> how -> why'' reasoning hierarchy, which is then consolidated into a final summary. Each stage addresses a distinct question:

\begin{enumerate}[label=\arabic*., leftmargin=*]
        
        \item \textbf{Literal Code Change Description (What):} This stage establishes an objective foundation by answering the question ``What precisely was changed?''. The model grounds its analysis in the raw syntax of the modification, preventing any deviation from the factual evidence.

        \item \textbf{Functional Impact Analysis (How):} This stage moves beyond syntax to answer ``How does this change achieve its effect?'' The model reasons about the operational consequences of the change, analyzing its direct impact on program behavior. This step serves as the crucial bridge between observing a code change and understanding its functional significance.

        \item \textbf{Change Category Inference (Why):} This stage addresses the motivational aspect behind the change by answering ``Why was this change made?'' Rather than focusing on syntactic patterns or functional consequences, the model shifts perspective to infer the developer’s high-level intent, such as bug fixing, feature implementation, or refactoring. It consolidates prior reasoning and guides the LLM to select a predefined intent category that best explains the intent behind the change. We utilize a predefined list, as these discrete categories are essential for the category filtering. An open-ended classification would be impractical, as it could introduce ambiguous synonyms (e.g., ``Bug Fix'' vs. ``Error Fix'') that would break this critical filtering step.
        
        \item \textbf{Intent Summary Synthesis:} This stage consolidates the outputs of the ``what -> how -> why'' stages. It distills the entire analytical chain into a coherent and concise narrative, suitable for a formal commit message.
    \end{enumerate}
\item \textbf{Structured Output:} The prompt mandates a structured format with \textbf{Intent Profile}: <Change Category, Intent Summary>.

\end{itemize}

By enriching each MCS with an \textbf{Intent Profile}, this stage provides the core semantic units for the final refinement stage.

\subsection{Refinement Stage: Grouper-Reviewer Collaborative Refinement Loop}
\label{sec:reasoning}
After inferring the intent of each change, the final challenge is to group them into coherent atomic commits. Traditional clustering algorithms perform this as a ``single-pass'' operation, \textbf{lacking any mechanism to verify or refine the logical correctness} of their output. To overcome this, we designed a collaborative stage that \textbf{emulates a realistic code review process}, featuring the \agentC and the \agentD in an iterative \textbf{grouper-reviewer collaborative refinement} loop.

\subsubsection{Phase 1: Intent-Driven Greedy Grouping}

The \agentC serves as the initial drafter, tasked with generating a plausible grouping of MCSs, denoted as $\mathcal{M}=\{m_1,\dots,m_n\}$. It operates with a \textbf{local and incremental perspective}, making placement decisions for each MCS based on the current state of group assignments. The goal is to produce a final set of coherent groups $\mathcal{G_\text{set}}$ using an \textbf{Intent-driven Greedy Grouping Algorithm} as shown in Algorithm~\ref{alg:intent_grouping}:

\begin{enumerate}[label=\arabic*., leftmargin=*]
    \item \textbf{Initialization}: The process begins by assigning the first MCS $m_1$ to a new group $G_1 = {m_1}$, which is added to the output set: $\mathcal{G_\text{set}} = {G_1}$. The Intent Profile of $m_1$ is used to initialize the group’s Representative Intent, serving as a semantic anchor for future comparisons.

    \item \textbf{Hierarchical Placement via Two-Stage Grouping}: For each subsequent MCS $m_i \in \mathcal{M}$, the \agentC determines its placement using a hierarchical two-stage strategy:
    \begin{itemize}[leftmargin=*]
    
        \item \textbf{Coarse-Grained Category Filtering:} It first filters $\mathcal{G_\text{set}}$ by selecting candidate groups whose Representative Intent shares the same \texttt{Change Category} as $m_i$. The resulting subset forms a candidate set $C_i \subseteq \mathcal{G_\text{set}}$.
        
        \item \textbf{Fine-Grained Intent Matching:} Among the candidates in $C_i$, the model performs semantic matching between the Intent Summary of $m_i$ and each group’s Representative Intent. This enables a more coherent grouping based on semantic closeness rather than label matching alone.
    \end{itemize}

    \item \textbf{Dynamic Intent Update}: If $m_i$ is assigned to an existing group $G_j \in C_i$, the \agentC immediately updates the group’s Representative Intent by synthesizing the semantic intent of $m_i$ with that of $G_j$. This dynamic adjustment ensures the group evolves to reflect its aggregated purpose. If no suitable group is found, a new group is created for $m_i$ and added to $\mathcal{G_\text{set}}$.
\end{enumerate}

This process yields a complete grouping proposal. However, because the \agentC's decisions are greedy and local, the final composition of a group may not be globally coherent. This is where the \agentD provides a necessary review.

\begin{algorithm}[!t]
\caption{Intent-Driven Greedy Grouping}
\label{alg:intent_grouping}

\KwData{A set of MCSs $\mathcal{M} = \{m_1, ..., m_n\}$, each with an Intent Profile.}
\KwResult{A set of groups $\mathcal{G}_{set}$.}

\SetKwFunction{FMain}{GreedyGrouping}
\SetKwFunction{FGetCat}{GetCategory}
\SetKwFunction{FGetRepRat}{GetRepIntent}
\SetKwFunction{FSetRepRat}{SetRepIntent}
\SetKwFunction{FLLMComp}{ComparativeJudgment}
\SetKwFunction{FLLMSynth}{SynthesizeIntent}
\SetKwFunction{FGetProfile}{GetIntentProfile}
\SetKwProg{Fn}{Function}{:}{}

\Fn{\FMain{$\mathcal{M}$}}{
    $\mathcal{G}_{set} \gets \varnothing$\;
    \If{$\mathcal{M}$ is not empty}{
        $G_1 \gets \{m_1\}$\;
        \FSetRepRat{$G_1$, \FGetProfile{$m_1$}}\;
        $\mathcal{G}_{set} \gets \{G_1\}$\;
        
        \For{each MCS $m_i$ from $m_2$ to $m_{|\mathcal{M}|}$}{
            $C_i \gets \varnothing$\;
            \For{each group $G_j$ in $\mathcal{G}_{set}$}{

                \If{\FGetCat{\FGetProfile{$m_i$}} == \FGetCat{\FGetRepRat{$G_j$}}}{
                    $C_i \gets C_i \cup \{G_j\}$\;
                }
            }
            
            $G_{best}$ $\gets$ \FLLMComp{$m_i$, $C_i$}\;
            
            \If{$G_{best}$ is a new group}{

                $G_{new} \gets \{m_i\}$\;
                \FSetRepRat{$G_{new}$, \FGetProfile{$m_i$}}\;
                $\mathcal{G}_{set} \gets \mathcal{G}_{set} \cup \{G_{new}\}$\;
            }
            \Else{
                $G_{best} \gets G_{best} \cup \{m_i\}$\;
                $new\_intent \gets \FLLMSynth(G_{best})$\;
                \FSetRepRat{$G_{best}, new\_intent$}\;
            }
        }
    }
    \KwRet{$\mathcal{G}_{set}$}\;
}
\end{algorithm}

\subsubsection{Phase 2: Review Coherence}
The \agentD acts as an expert reviewer, whose mechanism is designed to systematically emulate the cognitive process of a human expert. To establish a reliable benchmark for its judgment and avoid being biased by potential outliers, its review process is executed in two steps: 

\begin{enumerate}[label=(\arabic*), leftmargin=*]
    \item \textbf{Identify the Core Intent}: Human reviewers typically begin by understanding the overarching purpose behind a group of related code changes. Mirroring this behavior, the \agentD first identifies the \textbf{Largest Coherent Subset} within the proposed group, defined as the maximal subset of MCSs that can be jointly explained by a concrete development purpose. This inferred purpose is treated as the group’s dynamically established reference-level \textbf{core intent}, and serves as the basis for evaluating group coherence in the next step.

    \item \textbf{Detect the Outliers}: With the \textbf{core intent} established, the \agentD then examines each MCS in the group through qualitative reasoning by the LLM to assess whether its individual intent aligns logically with the shared core intent. We define an \textbf{outlier} as any MCS whose individual purpose is inconsistent with the established core intent and thus cannot be logically included in the \textbf{Largest Coherent Subset}. If no outliers are found, the group is marked with an \texttt{ACCEPT} decision. Otherwise, the group receives a \texttt{REJECT} decision, along with diagnostic feedback that includes both the identified outliers and the group's core intent.
\end{enumerate}

\subsubsection{Phase 3: Refine with Feedback}

This phase refines groupings based on feedback from Phase 2. Specifically, the \agentC receives a decision from the \agentD for each group: (1) an \texttt{ACCEPT}, which finalizes the group as valid; or (2) a \texttt{REJECT}, which indicates that a group $G_i$ contains one or more outliers $m_{\text{outlier}}$ that deviate from the group’s core intent. This decision serves as explicit feedback guiding further refinement. In response, the system activates a \textbf{smart correction and re-grouping protocol}.

\begin{enumerate}[label=(\arabic*), leftmargin=*]
    \item \textbf{Outlier Excision and Re-Grouping}: All identified outliers are first removed from the rejected group $G_i$. For each outlier, the \agentC seeks a more appropriate semantic context. 
    \begin{itemize}[left=0pt]
        \item Identify Candidate Groups: It filters the current set of groups to those sharing the same \textbf{Change Category} as the outlier.
        \item Semantic Judgment: The LLM evaluates whether the outlier semantically fits into any candidate group. If no suitable candidate group is found, the outlier is placed into a new singleton group.
    \end{itemize}

    \item \textbf{Resubmission for Validation}: The refined group $G_i$ (now without outliers) has its representative intent updated accordingly. All groups involved in this reorganization, including the revised $G_i$ and any new or modified groups, are then submitted again to the \agentD for another round of feedback.
\end{enumerate}

This feedback-driven loop of ``group-review-refine'' continues iteratively until all groups receive \texttt{ACCEPT} decisions, ensuring convergence to a set of semantically coherent and verifiably correct atomic commits.

\section{Experimental Setup}
\label{sec:experiments}

\begin{table*}[!t]
\rowcolors{0}{white}{white} 
\centering
\caption{Performance on the C\# dataset. All accuracy is presented as percentages (\%). Each cell displays results in the format \textbf{Avg / OA}, The C\# projects evaluated are CL (Commandline), CM (CommonMark), HF (Hangfire), HU (Humanizer), LE (Lean), NA (Nancy), NJ (Newtonsoft.Json), NI (Ninject), and RS (RestSharp). \toolname consistently outperforms all SOTA baselines across $\text{Acc}^c\%$, $\text{Acc}^a\%$, and both aggregated metrics (OA and Avg), with all improvements statistically significant ($p < 0.01$).}
\label{tab:RQ1_table}
\renewcommand{\arraystretch}{1.12}
\resizebox{0.92\textwidth}{!}{
\begin{tabular}{clcccccccccc}
\toprule
\multirow{2}{*}{\centering \textbf{Metrics}} & \multirow{2}{*}{\textbf{Baseline}} & \multicolumn{10}{c}{\textbf{Project (Avg\% / OA\%)}} \\
\cmidrule(l){3-12}
& & \textbf{CL} & \textbf{CM} & \textbf{HF} & \textbf{HU} & \textbf{LE} & \textbf{NA} & \textbf{NJ} & \textbf{NI} & \textbf{RS} & \textbf{Average} \\
\midrule

\multirow{8}{*}{$\text{Acc}^c\%$}
& \textit{Barnett et al.}~\cite{barnett2015helping} & 14.2/14.8 & 13.1/13.5 & 12.6/11.3 & 13.9/13.2 & 7.4/8.1 & 8.6/8.3 & 7.9/7.4 & 10.2/10.8 & 10.1/9.5 & 10.9/10.8 \\
& \textit{Herzig et al.}~\cite{herzig2013impact} & 28.1/28.7 & 27.5/27.2 & 28.8/28.1 & 27.3/27.9 & 27.4/29.2 & 28.8/29.5 & 28.3/28.6 & 26.9/26.1 & 31.5/31.7 & 28.3/28.6 \\
& \textit{$\delta$-PDG+CV}~\cite{parțachi2020flexeme} & 34.6/34.2 & 35.8/35.1 & 36.4/35.7 & 30.3/30.9 & 35.2/35.8 & 34.6/34.1 & 34.9/34.5 & 37.2/37.6 & 33.8/31.3 & 34.8/34.4 \\
& \textit{Flexeme}~\cite{parțachi2020flexeme} & 34.1/34.7 & 33.5/33.2 & 33.6/31.8 & 33.1/33.7 & 35.4/33.2 & 32.8/32.5 & 27.3/27.9 & 32.1/32.6 & 34.8/33.4 & 33.0/32.6 \\
& \textit{UTango}~\cite{li2022utango} & 46.5/46.1 & 45.7/45.3 & 46.9/45.6 & 44.2/44.8 & 46.4/45.1 & 44.7/44.3 & 41.9/41.5 & 46.2/46.8 & 46.4/43.1 & 45.4/44.7 \\
& \textit{HD-GNN}~\cite{fan2024detect} & 52.3/52.9 & 51.5/51.1 & 52.7/48.4 & 50.2/50.8 & 50.6/48.3 & 49.9/49.5 & 45.2/45.8 & 54.4/54.1 & 47.7/48.3 & 50.5/49.9 \\ \cmidrule(l){2-12}
& \toolname (Ours) & \textbf{58.4/58.7} & \textbf{57.2/57.5} & \textbf{58.6/58.1} & \textbf{56.8/56.3} & \textbf{56.4/56.9} & \textbf{56.5/56.1} & \textbf{53.7/53.2} & \textbf{60.4/60.8} & \textbf{54.6/55.1} & \textbf{57.0/57.0} \\
& Improvement$\uparrow$ & +6.1/+5.8 & +5.7/+6.4 & +5.9/+9.7 & +5.4/+5.5 & +5.9/+8.6 & +6.6/+6.6 & +7.5/+7.4 & +6.0/+6.7 & +6.9/+6.7 & +6.5/+7.1 \\
\midrule

\multirow{8}{*}{$\text{Acc}^a\%$}
& \textit{Barnett et al.}~\cite{barnett2015helping} & 20.9/19.2 & 20.1/20.6 & 15.5/15.9 & 25.3/18.8 & 16.7/18.1 & 9.4/9.8 & 13.2/15.6 & 14.8/14.3 & 13.5/12.9 & 16.6/16.1 \\
& \textit{Herzig et al.}~\cite{herzig2013impact} & 58.4/64.1 & 65.7/65.3 & 62.2/64.8 & 53.6/62.2 & 66.8/69.4 & 63.1/67.7 & 64.5/71.9 & 57.3/57.8 & 70.2/70.6 & 62.4/66.0 \\
& \textit{$\delta$-PDG+CV}~\cite{parțachi2020flexeme} & 81.3/80.8 & 90.5/90.1 & 86.7/87.3 & 69.9/69.4 & 78.2/84.8 & 83.6/86.2 & 78.8/82.4 & 94.1/94.7 & 64.6/70.1 & 80.9/82.9 \\
& \textit{Flexeme}~\cite{parțachi2020flexeme} & 87.5/82.1 & 70.8/70.3 & 77.2/79.8 & 70.6/81.2 & 80.8/80.4 & 87.1/84.7 & 62.6/71.2 & 80.9/80.5 & 86.3/82.9 & 78.2/79.2 \\
& \textit{UTango}~\cite{li2022utango} & 90.7/90.2 & 89.4/89.8 & 88.3/86.6 & 89.9/89.4 & 89.2/88.8 & 88.5/88.1 & 83.7/83.3 & 91.9/91.5 & 88.2/87.8 & 88.9/88.4 \\
& \textit{HD-GNN}~\cite{fan2024detect} & 91.6/91.1 & 90.8/90.4 & 90.3/89.9 & 90.7/90.2 & 91.5/90.1 & 89.8/89.4 & 86.3/86.9 & 93.7/93.2 & 90.5/89.1 & 90.6/90.0 \\ \cmidrule(l){2-12}
& \toolname (Ours) & \textbf{97.6/97.1} & \textbf{96.8/96.3} & \textbf{96.5/96.9} & \textbf{97.2/97.8} & \textbf{95.4/95.1} & \textbf{95.7/95.3} & \textbf{97.5/97.9} & \textbf{97.2/97.8} & \textbf{96.4/96.1} & \textbf{96.7/96.7} \\
&  \cellcolor{gray} Improvement$\uparrow$ & +6.0/+6.0 & +6.0/+5.9 & +6.2/+7.0 & +6.5/+7.6 & +3.9/+5.0 & +5.9/+5.9 & +11.2/+11.0 & +5.5/+4.6 & +5.9/+7.0 & +6.1/+6.7 \\
\bottomrule
\end{tabular}}
\end{table*}

This section presents various experiments to evaluate the effectiveness of \toolname, which aims to answer the following three Research Questions (RQs).
\begin{itemize}[left=0pt]
    \item \textbf{RQ1: How effective is \toolname on the common C\# Dataset and Java Dataset?}
    \item \textbf{RQ2: How effective is \toolname for complex commits?}
    \item \textbf{RQ3: How does each component contribute to the overall performance of \toolname?}
\end{itemize}

\subsection{Datasets}
In this work, we choose two synthetic datasets widely used in previous work:
(1) \textbf{C\# Dataset}~\cite{parțachi2020flexeme}: This dataset comprises 1,612 synthetic composite commits constructed from 9 open-source C\# projects. Each synthetic commit is created by amalgamating 2 to 3 atomic commits~\cite{parțachi2020flexeme, li2022utango, chen2022untangling}. (2) \textbf{Java Dataset}~\cite{li2022utango}: This dataset is more extensive, featuring over 14,000 synthetic composite commits sourced from 10 diverse open-source Java projects~\cite{li2022utango}. Each synthetic commit is formed by merging between 2 and 32 original atomic commits, offering a broad spectrum of tangling complexity.

\textit{Dataset Split}: We follow the chronological splitting methodology from \textit{UTango}~\cite{li2022utango}. For each project, commits are sorted by their creation date. For trainable baselines, we use the oldest 80\% of commits for training, the next 10\% for validation, and the most recent 10\% for testing. Non-trainable baselines and our approach are evaluated directly on the 10\% test set. 

\subsection{Baselines}
For the C\# dataset, we compared \toolname against six representative baselines supporting C\#. (1) \textit{Barnett et al.}~\cite{barnett2015helping}: A \textit{heuristic rule-based approach} considering def-use, use-use, and same-enclosing method relations among code changes. (2) \textit{Herzig et al.}~\cite{herzig2016impact}: A \textit{heuristic rule-based approach} combines various Confidence Voters and builds a triangle partition matrix to untangle the commits. (3) \textit{Flexeme}~\cite{parțachi2020flexeme}: A \textit{graph clustering-based} approach builds a $\delta$-NFG of commits and then applies agglomerative clustering to untangle the commits. (4) \textit{$\delta\text{-}PDG+CV$}~\cite{parțachi2020flexeme}: A variant of \textit{Flexeme} by applying \textit{Herzig et al.}’s confidence voters directly to $\delta$-PDG. (5) \textit{UTango}~\cite{li2022utango}: A \textit{graph clustering-based approach} builds a $\delta$-PDG from commits and then applies GNN and agglomerative clustering to untangle the commits. (6) \textit{HD-GNN}~\cite{fan2024detect}: A \textit{graph clustering-based approach} uses a hierarchical graph to detect hidden dependencies and then applies GNN to untangle the commits.

\begin{figure*}[!t]
  \centering
  \hspace{2.8cm}
  \includegraphics[width=0.75\linewidth]{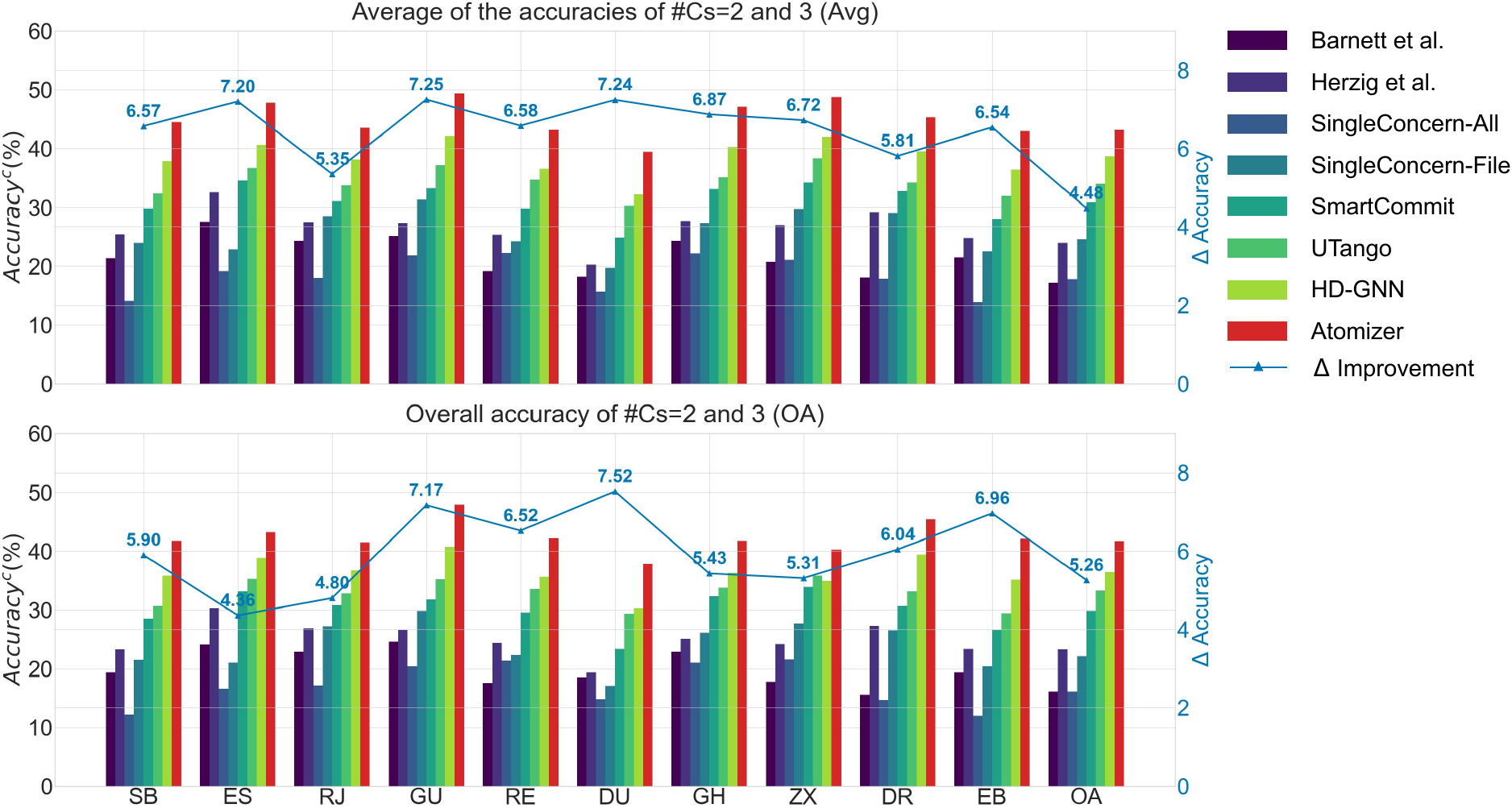} 
  \caption{Effectiveness on the Java Dataset. The Java projects evaluated are SB (Spring-boot), ES (Elasticsearch), RJ (RxJava), GU (Guava), RE (Retrofit), DU (Dubbo), GH (Ghidra), ZX (Zxing), DR (Druid), EB (EventBus).}
  \label{fig:RQ2.}
\end{figure*}

For the Java dataset, we compared \toolname against a diverse set of existing approaches that support Java, including rule-based methods: \textit{Barnett et al.}~\cite{barnett2015helping} and \textit{Herzig et al.}~\cite{herzig2016impact}, as well as three graph clustering-based methods: \textit{SmartCommit}~\cite{shen2021smartcommit}, \textit{UTANGO}~\cite{li2022utango}, and \textit{HD-GNN}~\cite{fan2024detect}. To ensure a more comprehensive comparison, we also included three additional strategies described in~\cite{shen2021smartcommit}:
(1) \textit{SingleConcern-All} is a simple rule-based approach that places all changes into one concern.
(2) \textit{SingleConcern-File} is a rule-based approach that puts the changes in each file into one concern.

\subsection{Evaluation Metrics}
To evaluate untangling accuracy, we adopt two widely used metrics from prior work~\cite{parțachi2020flexeme, li2022utango, chen2022untangling, fan2024detect}:

\begin{itemize}[left=0pt]

\item \textbf{$\text{Acc}_c$ (Changed Accuracy):} Introduced by \textit{UTango}~\cite{li2022utango}, this metric focuses only on the changed statements. It directly measures how accurately the model can group the actual code modifications that need to be untangled.

$$\text{Acc}_c = \frac{\# \text{Correctly predicted changed statements } }{\# \text{All changed statements}}$$

\item \textbf{$\text{Acc}_a$ (Absolute Accuracy):} Proposed by \textit{Flexeme}~\cite{parțachi2020flexeme}, this metric calculates the percentage of correctly classified statements among all statements (both changed and unchanged). It measures the model's overall correctness and its ability to distinguish changed code from the vast amount of unchanged code, without erroneously disturbing it. A high $\text{Acc}_a$ indicates that the model is not only grouping changes well but is also correctly leaving the stable parts of the files alone.

$$\text{Acc}_a = \frac{\# \text{Correctly predicted statements } }{\# \text{All statements in graph}}$$

\end{itemize}

To illustrate the different focus of these two metrics, suppose a commit contains 100 statements, with 95 unchanged and 5 changed. If a model correctly predicts all unchanged statements and 3 out of the 5 changed statements, then its core performance is $\text{Acc}_c = 3/5 = 60\%$, while its overall correctness is $\text{Acc}_a = (95 + 3)/(95 + 5) = 98\%$. Follow prior work~\cite{parțachi2020flexeme,li2022utango, chen2022untangling, fan2024detect}, we report both $\text{Acc}_a$ and $\text{Acc}_c$ for the C\# dataset, and only $\text{Acc}_c$ for the Java dataset. This decision is made to maintain consistency and ensure a fair and direct comparison with the methodologies established in the original benchmark papers for each respective dataset. We further report two aggregated accuracy metrics following prior work~\cite{parțachi2020flexeme, li2022utango, chen2022untangling, fan2024detect}: (1) \textbf{OA} (Overall Accuracy), computed by evaluating $\text{Acc}_c$ over all multi-concern commits in the test set collectively. It measures the model's raw performance across the entire dataset as a whole, treating every changed statement equally. (2) \textbf{Avg} (Average Accuracy), calculated as the average $\text{Acc}_c$ across commits with exactly two and three concerns, respectively. It measures the model's typical performance on a per-commit basis for the most common untangling scenarios, preventing a single large commit from skewing the evaluation.

\subsection{Implementation Details}
All experiments were run on a Linux server with an Intel i9-12900K CPU and dual RTX 4090 GPUs (24GB each). \toolname uses GPT-4o as the default reasoning agent, with decoding parameters set to temperature=0, top\_p=1, and n=1 for determinism. The iterative \textit{group-review-refine} loop runs up to 3 rounds to avoid infinite cycles. We also tested Gemini-2.5-Pro and DeepSeek-V3 under the same settings to assess model generalizability.

\section{Experimental Results}
\label{subsec:results}

\subsection{RQ1: Effectiveness on the C\# Dataset and Java Dataset.}

\paragraph{Effectiveness on the C\# Dataset} 
Table~\ref{tab:RQ1_table} shows the comparison on the C\# dataset. For the changed node prediction accuracy ($\text{Acc}^c\%$), \toolname significantly outperforms all state-of-the-art (SOTA) approaches. On average, \toolname surpasses \textit{Barnett et al.}, \textit{Herzig et al.}, \textit{Flexeme}, \textit{$\delta\text{-}PDG+CV$}, \textit{UTango}, \textit{HD-GNN} by 46.1\%, 28.7\%, 22.2\%, 24.0\%, 11.6\%, 6.5\% in terms of \textbf{Avg} accuracy, respectively. A similar trend is observed for the \textbf{OA} metric, with corresponding improvements of 46.2\%, 28.4\%, 22.6\%, 24.4\%, 12.3\%, 7.1\%. \toolname also achieves the highest overall performance on all-node accuracy ($\text{Acc}^a\%$). Notably, the results for all models are higher on $\text{Acc}^a\%$ than $\text{Acc}^c\%$ because they have correct classifications for the unchanged nodes by default. Notably, when compared to \textit{HD-GNN}~\cite{fan2024detect}, the strongest graph clustering-based baseline, \toolname yields an average gain of over 6.0\% across all three metrics. To validate the significance of the improvements, we conducted paired t-tests, which show that the performance gains of \toolname over all baselines are statistically significant, with all $p$-values falling below 0.01.

\paragraph{Effectiveness on the Java Dataset} 
Figure~\ref{fig:RQ2.} presents the results on the Java dataset. Across both \textbf{Avg} and \textbf{OA} metrics, \toolname consistently achieves the best performance. Compared to the strongest baseline \textit{HD-GNN}, \toolname obtains an average improvement of more than 5.5\%. Paired t-tests confirm that all improvements over the baselines are statistically significant, with \textit{p}-values below 0.01.

\find{\textbf{Answering RQ1:}
The experimental results demonstrate that \toolname consistently outperforms all SOTA baselines on both the C\# and Java datasets. On average, it surpasses the strongest baseline \textit{HD-GNN} by over 6.0\%$\uparrow$ and 5.5\%$\uparrow$, respectively. For both datasets, these improvements are statistically significant (\textit{p} < 0.01).}

\begin{table}[!t]
\centering
\caption{Performance comparison on C\# commits with varying complexity ($\text{Acc}^c\%$).}
\label{tab:rq2_complexity}
\renewcommand{\arraystretch}{1.1}
\resizebox{0.48\textwidth}{!}{
\begin{tabular}{ccccc}
\toprule
\textbf{Node Range} & \textbf{Number} & \textbf{\textit{HD-GNN}} & \textbf{\toolname (Ours)} & \textbf{Improvement} \\
\midrule
0-1000              & 605                    & 53.7                  & 59.6                    & +5.9\%               \\
1000-2000           & 385                    & 49.3                  & 58.2                    & +8.9\%               \\
2000-7000           & 513                    & 46.5                  & 58.5                    & +12.0\%              \\
>7000               & 109                    & 41.4                  & 57.6                    & +16.2\%              \\
\bottomrule
\end{tabular}}
\end{table}

\subsection{RQ2: Effectiveness for complex commits.}
To further investigate \toolname's performance under varying levels of complexity, we conducted a stratified analysis based on the size of the graph, specifically the number of nodes it contains. All commits in the C\# dataset were sorted by graph size and divided into four complexity groups: 0–1000, 1000–2000, 2000–7000, and >7000 nodes. Notably, graphs exceeding 7000 nodes are extremely large in the context of real-world version history~\cite{fan2024detect, chen2022untangling, li2022utango}; very few commits exceed this scale, with the average commit size typically under 1500 nodes. Thus, this group represents the most challenging cases for untangling. As Table~\ref{tab:rq2_complexity} shows, \textbf{as commit complexity increases, its performance remains stable or even improves slightly, while the accuracy of the state-of-the-art baseline \textit{HD-GNN} degrades sharply.} Specifically, \toolname maintains a stable accuracy of 58–59\% even for the largest commits, whereas \textit{HD-GNN} drops from 53.7\% to just 41.4\%. The accuracy gap between the two methods, therefore, grows from +5.9\% in the simplest group to a substantial +16.2\% in the most complex group. This divergence highlights the fundamental advantage of our method. \toolname is designed to preprocess to get focused MCSs and reason about developer intent, which allows it to handle large graphs robustly. In contrast, \textit{HD-GNN}, which relies on graph clustering, becomes increasingly vulnerable to the irrelevant dependencies in large commits, causing its performance to degrade significantly.

\find{\textbf{Answering RQ2:}
\toolname is particularly effective for complex commits with large graphs. As commit size increases, the performance of the strongest existing method \textit{HD-GNN} drops by over 12.3\%$\downarrow$, while \toolname maintains stable or even slightly improved accuracy.
}

\subsection{RQ3: How does each component contribute to the overall performance of \toolname?}

\begin{table}[!t]
\centering
\caption{Ablation study results on the C\# dataset ($\text{Acc}^c\%$).}
\label{tab:rq3_ablation}
\begin{tabular}{llcc}
\toprule
\textbf{Variant} &  \textbf{Avg Acc (\%)} & \textbf{OA Acc (\%)} \\
\midrule
\toolname                                         & 57.0                  & 57.0                 \\
\midrule
w/o Purification  & 48.1                  & 49.2                 \\
w/o RA-CoT                         & 49.8                  & 50.5                 \\
w/o Review                     & 51.2                  & 51.9                 \\
\bottomrule
\end{tabular}
\end{table}

To evaluate the performance of different components within the \toolname framework, we conducted a series of ablation studies, as shown in Table~\ref{tab:rq3_ablation}.
We designed three ablated variants of \toolname by disabling one key component at a time. These variants are compared against the full implementation:
\begin{itemize}[left=0pt]
    \item \textbf{w/o Purification}: Bypasses the Purification Stage, feeding raw, noisy file diffs directly to \agentB instead of cleaned MCSs. 
    \item \textbf{w/o IO-CoT}: Replaces the IO-CoT strategy in the Analysis Stage with a zero-shot prompt, requesting a direct intent summary from the LLM.
    \item \textbf{w/o Review}: Disables \agentD and the review-refine loop. The output is the \agentC’s initial greedy algorithm grouping without further validation.
\end{itemize}

The results in Table~\ref{tab:rq3_ablation} clearly demonstrate that each component makes a significant and positive contribution to the framework's overall performance.
A substantial performance degradation is observed in the \toolname w/o Purification and \toolname w/o IO-CoT variants. Removing the initial Purification Stage causes a sharp drop in accuracy, confirming our hypothesis that without distilling raw diffs into clean MCSs, the subsequent agents are confounded by structural noise, leading to flawed intent inference. Similarly, removing the IO-CoT strategy results in a major performance loss, which validates that explicitly guiding the LLM through a structured ``what -> how -> why'' reasoning process is essential for accurately discerning developer intent.
Finally, the \toolname w/o Review variant also shows a noticeable decline in accuracy. This result underscores the value of the Collaborative Grouping and Reviewing Stage. It proves that the holistic perspective of the \agentD is effective at identifying and correcting logical inconsistencies made during the \agentC's initial greedy grouping phase, thereby refining the output to a more coherent and correct state.

\find{\textbf{Answering RQ3:}
Each component of the \toolname framework makes a vital and positive contribution. The ablation study reveals that the Purification Stage is crucial for mitigating structural noise, the IO-CoT is essential for accurate semantic understanding, and the Collaborative Reviewing Stage provides a necessary validation mechanism that improves the final grouping accuracy. The results confirm that the synergy of these specialized stages is key to \toolname's excellent performance.
}

\section{Discussion}

\begin{table}[!t]
\centering
\caption{Performance of \toolname with different foundational LLMs on the C\# dataset ($\text{Acc}^c\%$).}
\label{tab:llm_comparison}
\begin{tabular}{lcc}
\toprule
\textbf{Foundational LLM} & \textbf{Avg Acc (\%)} & \textbf{OA Acc (\%)} \\
\midrule
GPT-4o (Primary)          & 57.0                  & 57.0                 \\
Gemini-2.5-Pro            & 56.9                  & 57.1                 \\
DeepSeek-V3               & 56.4                  & 56.8                 \\
\bottomrule
\end{tabular}
\end{table}

\subsection{Impact of \toolname on Different Foundational LLMs}
To validate the generalizability of the \toolname, we conducted an experiment where we substituted its core reasoning engine. While the primary results in this paper were achieved using GPT-4o, we replaced it with two other powerful Large Language Models: Gemini-2.5-Pro and DeepSeek-V3. The rest of the \toolname's architecture remained unchanged. All three models were configured with the same deterministic parameters (temperature=0, top\_p=1) and were evaluated on the C\# dataset to ensure a fair comparison. As shown in Table~\ref{tab:llm_comparison}, while GPT-4o yields the best performance, both Gemini-2.5-Pro and DeepSeek-V3 achieve competitive results. This confirms that \toolname's multi-agent design is robust, model-agnostic, and effective across different LLMs. This suggests that while a more powerful LLM can enhance the final accuracy, the \toolname architecture itself is the primary driver of its success.

\subsection{Evaluation on the Real-World Dataset}
\begin{table}[!t] 
\centering
\caption{Performance comparison on the expert-annotated MVD dataset~\cite{fan2024detect}.}
\label{tab:real_data_results}
\resizebox{0.48\textwidth}{!}{
\begin{tabular}{lccc} 
\toprule
\multirow{2}{*}{\textbf{Method}} & \multicolumn{2}{c}{\textbf{C\# MVD}} & \textbf{Java MVD} \\ \cmidrule(r){2-3} \cmidrule(l){4-4}
 & \textbf{Acc$^c$ (\%)} & \textbf{Acc$^a$ (\%)} & \textbf{Acc$^c$ (\%)} \\ \midrule
Barnett et al.~\cite{barnett2015helping} & 9.2  & 51.4 & 35.7 \\ 
UTango~\cite{li2022utango}               & 38.6 & 74.1 & 37.9 \\ 
HD-GNN~\cite{fan2024detect}              & 43.4 & 81.3 & 48.9 \\ 
\textbf{\toolname (ours)}                & \textbf{61.6} & \textbf{94.5} & \textbf{64.7} \\ 
\bottomrule
\end{tabular}}
\end{table}
To evaluate the performance of \toolname on real-world data, we conducted experiments on a small, custom dataset from~\cite{fan2024detect}, which is annotated by expert developers for detecting hidden dependencies in commits. As shown in Table~\ref{tab:real_data_results}, our results demonstrate that \toolname significantly outperforms strong baselines on this challenging dataset. Specifically, \toolname demonstrates a significant advantage in accuracy over other methods, achieving 61.6\% (Acc$^c$) on the C\# MVD task, 64.7\% on the Java MVD task, and a strong 94.5\% (Acc$^a$) on the real-world dataset. These results highlight that \toolname provides higher accuracy and stronger dependency detection capabilities when applied to real-world datasets.

\subsection{Cost Analysis}

We analyze the computational cost of \toolname, which correlates with task complexity (as shown in Table~\ref{tab:cost_analysis_transposed}). When processing commits, the average performance of GPT-4o varies significantly by dataset. For the C\# dataset (averaging 2–3 concerns), processing a commit takes approximately 48 seconds and consumes 16,500 tokens. However, for the more complex Java dataset (with a wider range of 2 to 32 concerns), this increases to an average of 63 seconds and 20,100 tokens. This same scaling trend in processing time and token usage is also observed with both Gemini-2.5-Pro and DeepSeek-V3. Based on mainstream API pricing, this results in an approximate cost per commit of \$0.053 (C\#) and \$0.064 (Java) using GPT-4o. Costs are lower with Gemini-2.5-Pro (\$0.047 / \$0.059) and DeepSeek-V3 (\$0.026 / \$0.032), respectively. The \agentB is the primary cost driver, accounting for 72\% of the total tokens, as it performs the detailed semantic analysis for each MCS. In complex scenarios involving commit graphs with over 7,000 nodes, Atomizer achieves an average improvement of about 16.2\% compared to traditional GNN-based methods, while maintaining a low cost of just \$0.069 per commit when using GPT-4o, demonstrating excellent cost-effectiveness. To further optimize cost, a possible approach is to combine low-cost traditional methods with powerful LLM-based methods. Specifically, for simple and unambiguous commits, the combined approach handles them using low-cost graph-based methods as an initial filter. For complex or semantically ambiguous commits, it adopts an LLM-based multi-agent framework. A key challenge will be designing an effective method for classifying the commit difficulty. This combined strategy should achieve a practical balance for real-world deployment. 

\begin{table}[t!]
\centering
\setlength{\tabcolsep}{12pt}
\caption{\textcolor{black}{Cost Analysis of \toolname.}}
\label{tab:cost_analysis_transposed}
\begin{tabular}{llrr}
\toprule
\multicolumn{1}{l}{\textbf{Model}} & \textbf{Metric} & \textbf{C\#} & \textbf{Java} \\
\midrule
\multirow{3}{*}{GPT-4o} &  Time (s) & 48 & 63 \\
&  Token & 16,500 & 20,100 \\
&  Cost (\$) & 0.053 & 0.064 \\
\midrule
\multirow{3}{*}{Gemini-2.5-Pro} &  Time (s) & 61 & 78 \\
&  Token & 15,300 & 18,470 \\
&  Cost (\$) & 0.047 & 0.059 \\
\midrule
\multirow{3}{*}{DeepSeek-V3} &  Time (s) & 55 & 69 \\
&  Token & 15,720 & 17,690 \\
&  Cost (\$) & 0.026 & 0.032 \\
\bottomrule
\end{tabular}
\end{table}

\subsection{Analysis of Frequency of Re-evaluation by the Reviewer Agent}
Statistical analysis of \toolname's re-evaluation frequency reveals that 85\% of C\# commits are completed in one round, 9\% in two, and 6\% in three. For the more complex Java dataset, this distribution shifts to 71\%, 17\%, and 12\%, respectively. Notably, 67\% of commits with over 16 concerns require three rounds. Furthermore, 80\% of re-evaluations create new singleton groups for ``outliers,'' while 20\% involve merging changes into existing groups.

\subsection{Threats to Validity}
\textbf{Internal Validity.} Despite using developer-labeled atomic commits as ground truth, inherent noise or minor entanglements may exist, potentially affecting absolute accuracy. Furthermore, while we fix the LLM temperature to 0 to minimize stochasticity, complete determinism is not guaranteed, and the \agentD remains limited by the underlying model's reasoning capabilities.

\noindent\textbf{External Validity.} Our findings, based on open-source C\# and Java datasets, may not fully generalize to proprietary systems or other programming languages. Additionally, while synthetic datasets (merged atomic commits) are standard, they may lack the unstructured complexity of real-world changes. We addressed this with a real-world dataset, though its limited scale and specific project domains may not cover all development scenarios.

\section{Conclusion}
In this work, we introduced \toolname, a multi-agent framework designed to untangle composite commits by capturing developer intent and providing a self-refinement mechanism. By employing an Intent-Oriented Chain-of-Thought (IO-CoT) strategy and an iterative ``review-and-refine'' loop, \toolname enables human-like self-correction. Experiments on C\# and Java datasets demonstrate that \toolname achieves state-of-the-art performance, significantly outperforming baselines, particularly on large, complex commits.

\begin{acks}
The authors greatly thank the anonymous reviewers for their constructive comments.
This work was supported by the National Natural Science Foundation of China (Grant No.62402506, No.62474196) and the Research Foundation from NUDT (Grant No. ZK24-05).
\end{acks}

\newpage

\balance
\bibliographystyle{ACM-Reference-Format}
\bibliography{references}

\end{document}